\documentclass{elsart}

\usepackage{graphicx}
\usepackage{amssymb}

\usepackage[dvipsnames]{color}

\begin{document}

\begin{frontmatter}



\newcommand{\3}{\ss}
\newcommand{\n}{\noindent}
\newcommand{\eps}{\varepsilon}
\newcommand{\be}{\begin{equation}}
\newcommand{\ee}{\end{equation}}
\newcommand\ltsima{$\; \buildrel &lt; \over \sim \;$}
\newcommand\simlt{\lower.5ex\hbox{\ltsima}}
\newcommand\gtsima{$\; \buildrel &gt; \over \sim \;$}
\newcommand\simgt{\lower.5ex\hbox{\gtsima}}

\title{Acceleration of dust particles by low-frequency Alfv\'en 
waves }

\author[IF]{V.~~Prudskikh}\ead{E-mail:slavadhb@mail.ru},~
\author[DP]{Yu.~A.~Shchekinov}\ead{E-mail:yus@phys.rsu.ru}

\address[IF]{Institute of Physics, South Federal University, 
Stachki St. 194}

\address[DP]{Department of Physics, South Federal University,
Sorge St. 5, Rostov-on-Don, 344090 Russia}

\begin{abstract}
We investigate the efficiency of acceleration of charged dust 
particles by low-frequency Alfv\'en waves in nonlinear approximation. 
We show that the longitudinal acceleration of dust particles 
is proportional to the square of the soliton amplitude $O(|b_m|^2)$, 
while the transversal acceleration is of $O(|b_m|)$. In the conditions 
of the interstellar medium the resulting velocity of dust particles 
can reach 0.3 to 1 km s$^{-1}$. 

\end{abstract}

\begin{keyword}


\end{keyword}

\end{frontmatter}

\section{Introduction}
The presence of charged dust grains in space plasma may play 
crucial influence on the overall dynamics, as for instance, the 
breaking down of frozen-in-field conditions \cite{Kamaya00} and 
dissipation of magnetic flux \cite{nakano02} in the interstellar 
molecular clouds. The dynamical importance of dust grains, and 
circulation of their mass (dust destruction and growth) in the  interstellar medium is essentially determined by their kinetic 
temperature, and therefore understanding of how they are 
accelerated is of significant interest \cite{lazar,yan}. In general, 
magnetohydrodynamic (MHD) waves are known as an efficient heating 
source of plasma components. For instance, interaction of dispersive 
Alfv\'en waves with plasma results in Joule heating of the electrons, 
generation of zonal flows and formation of localized Alfv\'en structures 
and Alfv\'en vortices \cite{shuk04}. In a dusty plasma 
MHD waves, in particular, compressible fast 
MHD modes, are found to be efficient in heating of translational 
degrees of freedom of dust particles through gyro-resonant 
acceleration \cite{lazar}. More recently  
ponderomotive forces of the shear Alfv\'en waves are shown to 
be a powerfull mechanism of dust acceleration \cite{shuk03}, such 
that in a 
short time a microsized dust particle of mass $m_d\sim 10^{-12}$ g 
can reach a few km s$^{-1}$ velocity -- comparable to the phase 
velocity of the Alfv\'en wave. In \cite{shuk03} a steady nonuniform 
field envelope with a fixed field gradient (caused, for instance, by 
a nonuniform plasma density distribution along the wave vector) is 
described. Contrary, in our analyses below the ponderomotive force is 
unsteady in the rest reference frame where the pulse envelope moves with the group velocity. In this approach we are able to derive the 
perturbation amplitudes in the reference frame connected with the 
pulse, for densities and longitudinal velocities of 
plasma components driven by the mutual action of Coulomb and 
ponderomotive forces. Therefore, the nonuniformity of the field 
amplitude and the corresponding plasma acceleration in the pulse follows 
in our approach from a self-consistent solution of MHD equations.

In \cite{shuk03} they considered a 
relatively high-frequency limit, corresponding to frequencies much 
higher than the dust cyclotron frequency $\omega_d$. It might be 
expected though that in the low-frequency end $\omega\ll \omega_d$ 
the presence of dust can have an important influence on dynamics of 
Alfv\'en waves. In particular, one can think that if dust particles are accelerated, the energy gained by them from the 
Alfv\'en wave can be sufficient to decrease its amplitude, 
which in turn can reduce the acceleration efficiency. In this paper 
we explore the efficiency of acceleration of dust grains by the 
nonlinear low-frequency Alfv\'en waves, and show that the nonlinearity strongly 
restricts the ability of Alfv\'en waves to transfer their energy to 
charged dust grains.  Such solitons can be actually treated as elementary objects of MHD turbulence in the low-frequency range. We 
follow therefore recent discussion of interaction of such low-frequency 
Alfv\'en solitons with the plasma \cite{cram,haq}.

In the conditions of the interstellar plasma 
this low-frequency range, $\omega\ll \omega_d\sim 3\times 10^{-11}$ Hz, 
can be of great interest from the point of view of dust acceleration 
because a considerable energy fraction of the interstellar turbulence 
can be locked in low-frequency motions \cite{arm}. Of course, 
nonlinear interactions move low-frequency Alfv\'en perturbations 
toward higher frequences. However, it seems natural to assume that 
in the interstellar environment a steady state spectrum of perturbations 
establishes on a relatively short time scale (definitely shorter than 
100 Myr -- the crossing time for a spiral wave), so that perturbations 
are always present in the low-frequency end.

We proceed as follows. In Section 2 we find the dispersion relation and 
the group velocities for longitudinal Alfv\'en and fast MHD waves, 
in Section 3 we derive the nonlinear wave 
equation; Section 4 contains estimates 
of the velocities of dust grains accelerated by the Alfv\'en 
and fast MHD soliton waves; summary of the results is given in Section 5.

\section{Group velocity}

In a fully ionized dusty plasma the dispersion relation for 
longitudinal Alfv\'en and fast MHD modes can be 
written as \cite{mihal}

\begin{equation}
\label{disp1}
N^2=\epsilon_1\pm \epsilon_2, 
\end{equation} 
where $N=kc/\omega$, 

\begin{equation}
\epsilon_1=1-\sum\limits_{\alpha}^{}{\omega_{p,\alpha}^2\over 
\omega^2-\omega_{\alpha}^2},~\epsilon_2=-\sum\limits_{\alpha}^{}
{\omega_{p\alpha}^2\omega_\alpha\over \omega(\omega^2-\omega_\alpha^2)},
\end{equation} 
$\alpha=e,~i,~d$, $\omega_{p\alpha}$ and $\omega_\alpha$ are the 
plasma and cyclotron frequencies of the $\alpha$th component. In the  low-frequency 
limit $\omega\ll \omega_d$ equation (\ref{disp1}) can be written in the form 

\begin{equation}
\label{d1}
{\omega^2\over k^2}=v_A^2\left[1+{\rho_d\over \rho_i}\pm
{\omega\over \omega_i}\left(1-{m_d\rho_d\over Z_dm_i\rho_i}\right)
\right]^{-1},
\end{equation} 
where signs $\pm$ belong to the Alfv\'en and fast MHD waves, 
respectively, $v_{Ai}=B_0/\sqrt{4\pi\rho_i}$ is the Alfv\'en speed. 

Assuming $m_d\rho_d/Z_dm_i\rho_i\gg 1$ and 
$\omega\ll \rho_d\omega_d/\rho$, $\rho=\rho_i+\rho_d$, 
one can get from here 

\begin{equation}
\omega^2=k^2v_A^2\left(1\pm{\rho_d\over \rho}kr_{Ad}\right),
\end{equation}
where $v_A^2=B_0^2/4\pi\rho$, $r_{Ad}=v_A/\omega_d$. The group 
velocity and its derivative for Alfv\'en waves are 

\begin{equation}
\label{gra}
u={d\omega\over dk}=v_A\left(1+{\rho_d\over \rho}r_{Ad}k\right),~ 
{du\over dk}={\rho_d\over \rho}v_Ar_{Ad}>0,
\end{equation}
and for fast MHD mode 

\begin{equation}
\label{grbmz}
u={d\omega\over dk}=v_A\left(1-{\rho_d\over \rho}r_{Ad}k\right),~ 
{du\over dk}=-{\rho_d\over \rho}v_Ar_{Ad}<0. 
\end{equation}

\section{Nonlinear parabolic equation for low-frequency Alfv\'en 
and fast MHD waves}

Following the procedure described in \cite{shuk05,shuk06} one can derive 
the nonlinear Schr\"odinger equation for the envelope of the 
Alfv\'en wave package in the form 
\begin{equation}\label{schr}
\pm i{\partial E_1\over \partial \tau}+{u'\over 2} 
{\partial^2E_1\over \partial\xi^2}-{u\over 2kc^2}\sum\limits_{k}^{}
{\omega_{pk}^2\omega\over \omega+\omega_k}
\left(N_k-{kv_{zk}\over \omega}{\omega_k\over \omega+\omega_k}
\right)E_1=0,
\end{equation} 
where 
\begin{equation}
u'=\pm{\rho_d\over \rho_i+\rho_d}r_{AD}v_A,
\end{equation} 
is the dispersion of the group velocity; sign ``plus'' belongs to the 
left-polarized, ``minus'' to the righ-polarized waves, $N_k=\tilde n_k/
n_{k0}$ is the normalized density perturbation of $k$th component. In the 
left-polarized wave $\omega_k$ is positive for the electrons and dust 
particles, and negative for the ions, in the right-polarized Alfv\'en 
mode $\omega_k$ is positive for the ions and negative for the electrons and dust particles.

Assuming $v_{Ti}\ll v_{A}\ll v_{Te}$ one can neglect the 
electron inertia and write the equation of motion in the form 
\begin{equation}\label{eeq}
{\partial N_e\over \partial z}={e\over T_e}{\partial\phi\over \partial z} 
-{e^2\over M\omega_i\omega}\left({\partial\over \partial z}+{k\over \omega} {\partial\over\partial t}\right){|E|^2\over T_e}.
\end{equation}
The equations of longitudinal motion of the ions and dust are 
\begin{equation}\label{ieq}
{\partial V_z\over\partial t}=-{e\over M}{\partial\phi\over\partial z}
+{e^2\over M^2\omega(\omega_i-\omega)}\left({\partial\over\partial z}+
{k\over\omega}{\omega_i\over\omega_i-\omega}{\partial\over\partial t}
\right)|E|^2,
\end{equation}
\begin{equation}\label{deq}
{\partial w_z\over\partial t}=-{Ze\over m_d}{\partial\phi\over\partial z}
+{Z^2e^2\over m_d^2\omega(\omega_d+\omega)}\left({\partial\over\partial z}+
{k\over\omega}{\omega_d\over\omega_d+\omega}{\partial\over\partial t}
\right)|E|^2.
\end{equation}
Last terms in the r.h.s. of equations (\ref{eeq})-(\ref{deq}) describe 
the ponderomotive force acting on the particles by the moving wave. 

With using expansion over a small parameter $\omega/\omega_i\ll 1$ one can transform equation of motion of the ions and dust (\ref{ieq})-(\ref{deq})  in the reference frame moving with the wave group velocity to the form 
\begin{equation}\label{rieq}
-u{\partial V_z\over\partial\xi}=-{e\over M}{\partial\phi\over\partial\xi}+{f+F\over M},
\end{equation}
\begin{equation}\label{rdeq}
-u{\partial w_z\over\partial\xi}=-{e\over m_e}{\partial\phi\over\partial\xi}-{Zf-m_dF/M\over m_d},
\end{equation}
where
\begin{eqnarray}
f={\partial W_1\over\partial\xi}={e^2\over M\omega\omega_i}
\left(1-{ku\over\omega}\right){\partial|E|^2\over\partial\xi},\nonumber\\
F={\partial W_2\over\partial\xi}={e^2\over M\omega_i^2}
\left(1-{2ku\over\omega}\right){\partial|E|^2\over\partial\xi}.\nonumber
\end{eqnarray}

After integration of (\ref{eeq}) and (\ref{rieq})-(\ref{rdeq}) one can 
find the density perturbations as 
\begin{eqnarray}
&&N_e={e\phi-W_1\over T_e},~N_i={\e\phi-W_1-W_2\over Mu^2},\nonumber\\ 
&&N_d=
{-Ze\phi+ZW_1-m_dW_2/M\over m_du^2}.\nonumber
\end{eqnarray} 
Then the quasineutrality condition 
\begin{equation}\label{quasi}
n_{i0}N_i=n_{e0}N_e+Zn_{d0}N_d,
\end{equation}
gives the following solution for the Coulomb potential 
\begin{equation}
{e\phi\over T_e}={W_1\over T_e}-{W_2\over M(u^2-c_s^2)},
\end{equation}
and the explicit solutions for the density perturbations 
\begin{equation}\label{dens}
N_e=-{W_2\over M(u^2-c_s^2)},~N_i=-{W_2\over Mu^2}\left(1+{T_e/M\over 
u^2-c_s^2}\right),~N_d\simeq-{W_2\over Mu^2},
\end{equation}
here $c_s^2n_{i0}T_e/n_{e0}M$ is the square of the ion-sound dust speed. 
The longitudinal velocities are connected through the continuity equations 
with the density perturbations by the equations 
\begin{equation}\label{vels}
v_{zk}=uN_k.
\end{equation}

With accounting of equations (\ref{dens}) and (\ref{vels}) in the last terms in equation (\ref{schr}) one can reduce it to the following form 
for the new variable $b=B/B_0=kcE/\omega B_0$ 
\begin{equation}\label{nonl2}
\pm i{\partial b\over\partial\tau}+{u'\over 2}{\partial^2b\over\partial
\xi^2}-{1\over 2}kv_A{u^2-\rho_dc_s^2/\rho\over u^2-c_s^2}|b|^2b=0.
\end{equation}

The Lighthill criterion for the modulational instability 
fulfils when the signs of $u'$ and of the nonlinear term in (\ref{nonl2}) 
coincide. Thus, for Alfv\'en wave with $u'>0$ modulational unstability 
takes place whenever the conditions $c_s^2\rho_d/\rho<u^2<c_s^2$ hold. 
In the interstellar plasma with $\rho_d/\rho\sim 0.01$ both these 
conditions are valid. 
For the right-polarized Alfv\'en waves with $u'<0$ 
modulational instability can develop only 
in one of the two cases: either $u^2>c_s^2$ or $u^2<c_s^2\rho_d/\rho$, 
which hardly can hold in the conditions of the interstellar matter. 
Therefore, left-polarized Alfv\'en waves with the amplitude higher 
than a critical value experience in 
the interstellar plasma modulational instability which results in formation 
of a sequence of Alfv\'en solitons. 
In the limit $u\to c_s$ the free term in (\ref{nonl2}) becomes singular, 
which means that the quasineutrality condition is not 
valid anymore.

Using the standard procedure we represent the magnetic field as $b=b_\perp 
e^{i\Phi}$, where $b_\perp=b_\perp(\zeta-vt)$, $\Phi=\Phi(\zeta-v_1t)$, 
what means that the envelope solution we are seeking for  
drifts slowly (i.e. $v\ll u$) in the reference frame moving with the 
group velocity, while the phase $\Phi$ drifts with the velocity 
$v_1\ll u$. Separating the real and imaginary parts of (\ref{nonl2}) 
we arrive at 

\begin{eqnarray}
\label{nonl3}
&&{\partial^2 b_\perp\over\partial \zeta^2}-
{v(v-2v_1)\over u'^2}b_\perp+{2\alpha\over u'}b_\perp^3=0,
\nonumber\\
&&{\partial\Phi\over\partial\zeta}={v\over u'},
\end{eqnarray}
where 

\begin{equation}
\alpha=-{\omega\over 4}{u^2-c_s^2\rho_d/\rho\over u^2-c_s^2}.
\end{equation}
The solution of (\ref{nonl3}) is 

\begin{equation}
\label{sol}
b_\perp=b_m{\rm ch}^{-1}\left({\zeta-vt\over \Lambda}\right), 
\end{equation}
where $\Lambda=\sqrt{u'/\alpha b_m^2}$, and $v_1$ can be found as 
$v_1=v(1-\alpha u'b_m^2/v^2)/2$. In general, the solution (\ref{sol}) 
is determined by the two free parametes: $b_m$ and $v$. It is readily seen 
that collisions of dust particles with ions and atoms are unimportant on 
the scales of interest. Indeed, the characteristic length of the soliton 
(\ref{sol}) for $\omega\ll \omega_d$ can be estimated as $\Delta\zeta\gg 
r_{Ad}b_m^{-1}$, or for typical parameters in the interstellar plasma 
$\Delta\zeta\gg 10^{15}b_m^{-1}$ cm, which on the lower end is much less 
than the drag free path of dust particles: $\ell\sim 3\times 10^{19}n^{-1}$ cm, where the mass ratio $m_d/m\sim 10^{10}$ and mean 
grain radius $a=0.1\mu$m are assumed. 

\section{Longitudinal and transversal acceleration of dust particles}

In the laboratory reference frame longitudinal acceleration of dust 
particles is determined by the last equation in (\ref{dens}) and (\ref{vels}): 
$|V_{dz}|\simeq u|b|^2/2$, so that within the validity of the approximation 
the amplitude of dust velocity on nonlinear Alfv\'en wave can be of 
several percents of the Alfv\'en speed, which in the conditions of the 
interstellar plasma can vary from $v_A\sim 3$ km s$^{-1}$ in molecular 
gas ($\rho\sim 10^{-22}$ g cm$^{-3}$, $B\sim 10\mu$G) 
to $\sim 10$ km s$^{-1}$ in diffuse HI phasei ($\rho\sim 10^{-24}$ g cm$^{-3}$, 
$B\sim 3\mu$G). Therefore, reasonable 
conservative estimate of dust acceleration along magnetic lines 
in such environments may be of $0.3$ to $1$ km s$^{-1}$, respectively, 
which however depends on exact value of the soliton amplitude. 
This is much lower than the estimate obtained in \cite{shuk03}. 
Similarly, the dust velocity in perpendicular 
direction due to acceleration by the soliton can be estimated as $|V_{d\perp}|\simeq \omega |b_\perp|/k\
\sim u|b|$, which is therefore a factor of $b_m^{-1}$ larger than 
acceleration in the longitudinal direction. 

The acceleration of dust grains by a regular low-frequency Alfv\'en 
soliton ($\omega\ll \omega_d$) is connected with the ponderomotive force as in 
\cite{shuk03}, and differs qualitatively from the 
mechanism described in \cite{lazar}, where translational heating 
of dust particles is associated with stochastic acceleration of dust particles by MHD waves with $\omega\sim \omega_d$ through 
cyclotron resonant interactions.
Indeed, it is readily seen that dust particle moves 
under the action of soliton wave only within the soliton, while it turns 
to the rest outside the wave, e.i. at $\zeta\to -\infty$, and the only result of the action of the soliton is a shift of a particle by 
$\Delta \zeta\sim \Lambda$. 

The analysis of possible turbulent acceleration of dust particles by a random set 
of low-frequence Alfv\'en solitons is out of the scape of our paper. One can speculate only 
that as far as translational 
heating of dust particles by low-frequency MHD waves is concerned, 
one can connect it with an ensemble of solitons passing randomly through 
a given volume of the interstellar space. In this picture  
the effective  translational temperature is determined by the spectrum 
of low-frequency MHD solitons.

 
\section{Summary}

In this paper we considered acceleration of charged dust particles by nonlinear 
low-frequency MHD waves under the conditions typical for 
interstellar plasma. We have shown that contrary to the case of 
intermediate frequencies $\omega_d\ll\omega\ll\omega_i$ described in 
\cite{shuk03}, the acceleration efficiency is strongly limited by the 
amplitude of the nonlinear wave: while transversal velocities of dust 
particles $V_{d\perp}$ are proportional to the amplitude of the soliton, 
the longitudinal velocities $V_{dz}$ are of the second order of the 
amplitude. As a consequence, dust particles can be accelerated by such 
low-frequency waves to the velocities of few to tens percents of the 
Alfv\'en speed.  

This work was supported by the Federal Agency of Education (project code 
RNP 2.1.1.3483), by RFBR (project codes 05-02-17070 and 06-02-16819), and 
by the German Science Foundation (DFG) within Sonderforschungsbereich 591.

\newpage



\end{document}